\documentstyle[prl,aps,psfig]{revtex}  
\begin{document}
\draft

\twocolumn  
\title{Anomalous transport in normal-superconducting and 
ferromagnetic-superconducting nanostructures}

\author{R. Seviour$^*$, C. J. Lambert$^*$ and A. F. Volkov$^{\dagger*}$}
\address{$^*$ School of Physics and Chemistry,
Lancaster University, Lancaster LA1 4YB, U.K.\\
$^{\dagger}$Institute of Radioengineering and Electronics of the Russian
   Academy of Sciencies, Mokhovaya str.11, Moscow
   103907, Russia.}

\date{\today}
\maketitle

\begin{abstract}

 We have calculated the temperature dependence of the conductance variation 
($\delta S(T)$) of mesoscopic superconductor normal metal(S/N) structures, 
in the diffusive regime, analysing 
both weak and strong proximity effects. We show that in the 
case of a weak proximity effect there are two peaks in the dependence of $\delta S(T)$ 
on temperature. One of them (known from previous studies) corresponds to a 
temperature $T_1$ of order of the Thouless energy ($\epsilon_{Th}$), and another,
 newly predicted maximum, occurs at a temperature $T_2$ where the energy gap in 
the superconductor $\Delta(T_2)$ 
is of order $\epsilon_{Th}$. In the limit $L_{\phi}<L$ the temperature $T_1$ is 
determined by $D \hbar /L^2_{\phi}$ ($L_{\phi}$ is the phase breaking length), 
and not $\epsilon_{Th}$. We have also calculated the voltage dependence $
\delta S(V)$ for a S/F structure (F is a ferromagnet) and predict non-monotonic 
behaviour at voltages of order the Zeeman splitting.
\end{abstract}


 Since the late 1970's it has been known that the conductance ($G$) of S/N 
mesoscopic structures depends on temperature ($T$) (and voltage ($V$)) in a 
non-monotonic way (see reviews \cite{r1,r2}). This behaviour was first predicted 
in Ref. \cite{r3} where a simple point S/N contact was analysed. The authors of 
Ref. \cite{r3}, using a microscopic theory and assuming that the 
energy gap in the superconductor ($\Delta$) is much less than the Thouless energy 
$\epsilon_{Th} \equiv \hbar D/L^{2}$ ($D$ is the diffusion constant),  
showed that the zero-bias conductance $G$ coincides at zero temperature with 
its normal state value ($G_{n}$). With increasing $T$, $G$ exhibits a 
non-monotonic behaviour, 
increasing to a maximum of $G_{max} \approx 1.25 G_{n}$ at $T_m \approx \Delta(T_m)$ 
and then decreasing to $G_{n}$ for $T>T_m$.

 Recently mesoscopic S/N structures have been fabricated in which the limit 
$\Delta >> \epsilon_{Th}$ is realised. In this case Nazarov and 
Stoof \cite{r4} (also see \cite{r5,r6,r19}) argued that the temperature dependence of 
the conductance $G$ has a similar non-monotonic behaviour with a maximum at a 
temperature comparable with the Thouless energy, while simultaniously Volkov, 
Allsopp and Lambert \cite{r7} predicted that the voltage dependence of the conductance 
in an S/N  mesoscopic structure (Andreev interferometer) has a similar form 
with a maximum at $eV_{m} \approx \epsilon_{Th}$. This non-monotonic behaviour  
has been observed both in very short S/N contacts 
\cite{r8} and in longer mesoscopic S/N structures \cite{r9,r10,r11,r22}. 
In ref \cite{r6} it was noted  that the conductance $\delta G = G - G_{n}$ 
consists of two contributions. The first, $\delta G_{DOS}$, is negative due to 
a proximity effect induced decrease in the density of states (DOS) of the normal 
wire which makes contact with a superconducting strip \cite{r12}. The other 
contribution $\delta G_{MT}$ (positive) is 
analogous to the Maki-Thompson (MT) contribution to the paraconductivity 
 of S/N/S and N/S/N mesoscopic structures and was calculated 
in \cite{r13}. At $T=0$ and $V=0$ both contributions to the conductance 
are equal, as $T$ or $V$ increase the contribution $\delta G_{MT}$ dominates 
until a maximum is reached, then both these contributions decay. 

 During the past decade a great deal of interest in the transport properties of 
N/S nanostructures has originated from a desire to use superconductivity as a 
probe into phase-coherent transport. Indeed the above experiments reveal little 
about the superconductor itself, since the non-monotonic behaviour occurs at an energy 
much lower than $\Delta(0)$. In this 
paper we calculate the conductance of mesoscopic S/N structures (see Fig. 1) 
over a wide temperature range ($0<T \leq T_{c}$), and show that in the 
dependence of $\delta G(T)$, a second maximum may appear near $T_c$ when 
$\Delta(T_m)$ is of order the Thouless energy. Consequently, in contrast with 
 non-monotonic phenomena studied to-date, this peak provides a novel quasi-particle 
transport probe for the energy gap of the superconductor. Indeed  we show later 
that this second 
maximum is very sensitive to the damping rate inside the superconductor.We also show that 
if the depairing rate $\gamma$ (for example, due to magnetic impurities) 
is not small 
compared to $\epsilon_{Th}$ then the maxima in $\delta G(T)$ occur at a 
temperature $T_1\approx \gamma$ and $\Delta(T_1) \approx 
\gamma$. 

 We consider S/N mesoscopic structures of the form shown in Fig. 1A 
and 1B. Although they 
differ slightly from each other, in the limit $l_1<<1$  ($l_1=L_1/L$)
the formulae for the conductances of these systems are identical. We assume the
 metals are diffusive and employ the well developed quasiclassical Green's 
function technique (see for example \cite{r14}) which has been widely used for 
studying transport phenomena in S/N mesoscopic structures 
\cite{r2,r3,r4,r5,r6,r7,r15,r16,r17,r18,r19,r21,r22}. 
Using the Keldysh formalism, the 
conductance variation of the structure shown in Fig. 1B is given by \cite{r16}, 

\begin{equation}
\delta S=\int^{\infty}_0 d\epsilon \beta F_{v}'(\epsilon) \frac{t(\epsilon)}
{1-t(\epsilon)}
\label{eq1}
\end{equation}

 where all the quantites are dimensionless (dimensional quantities will be 
denoted by a tilde); $\delta S=(G-G_n)/G_n, \beta=(2T)^{-1}, F_{v}'(\epsilon)= 
[\cosh^{-2} \left(\epsilon +eV\right)\beta$  $+\cosh^{-2}  \left(\epsilon-eV
\right)\beta ]/2$, $t(\epsilon)=<\tanh^{2} \left(Re (u(\epsilon,x)\right))>,
<A>=(1-l_1)^{-1} \int^{1}_{l_1} dx A$. All energies and voltages are measured 
in units of the Thouless energy ($\epsilon_{Th}$), the function $u(\epsilon,x)$ 
is related to the condensate and normal Green's functions: $F^{R(A)}=\sinh
(u^{R(A)}), G^{R(A)}=\cosh(u^{R(A)})$, which obey the Usadel equation, 

\begin{equation}
\partial^{2}_{xx} u^{R(A)}-(k^{R(A)})^{2}\sinh u^{R(A)}=0
\label{eq2}
\end{equation}

 where $(k^{R(A)})^{2}=\gamma\mp 2 i \epsilon$. Eq. (\ref{eq2}) can be solved 
numerically \cite{r4,r5}, and in some limiting cases analytically 
\cite{r3,r4,r5,r6,r7,r15,r16,r17,r18,r19,r21,r22}. 

 First we consider the simplest case, where the proximity effect is weak, 
this occurs when the condensate function in the N wire $F^{R(A)}$ is small 
($|F^{R(A)}|\approx u^{R(A)} <<1$) \cite{r6,r7}. In this limit Eqs. (\ref{eq1}) 
and (\ref{eq2}) can be linearized. Also if the length $L$ is longer than the 
phase breaking length (if $\gamma>>1$) then the N wire may be considered as 
infinitely long, thus the solution to Eq.(\ref{eq2}) is determined by the 
expression \cite{r17}, 

\begin{equation}
\tanh(u/4)=\tanh(u_0/4) e^{-kx} 
\label{eq3}
\end{equation}

 where $u_0$ is determined by the boundary condition at the S/N interface. In the 
case of a good contact at the $N/N'$ interface the function $u$ should be zero 
at this interface ($x=1$), whereas Eq. (\ref{eq3}) gives a nonzero value for 
$u(\epsilon,1)$. The correction to the solution (\ref{eq3}) is small provided 
that $\gamma>>1$. In the case of the weak proximity effect we solve the 
linearized form of Eq (\ref{eq1}) with the boundary condition at the S/N interface 
\cite{r2},

\begin{equation}
r_b \partial_x \hat{F}^{R(A)}= -\hat{F}^{R(A)}_s|_{x=0} 
\label{eq4}
\end{equation}

 where $r_b$ is the ratio of the S/N interface resistance to the resistance of 
the N wire of length $L$; $\hat{F}^{R(A)}_s= i \hat{\sigma}_y F^{R(A)}_s \cos
(\phi/2)$ is the condensate Green's function in the superconductor, $F^{R(A)
}_s= \Delta/\sqrt{(\epsilon\pm i \Gamma)^2 -\Delta^2}$, $\Gamma$ is the damping 
rate in the spectrum of the superconductor and $\phi$ is the phase difference 
between the superconductors. In the case of zero $N/N'$ interface resistance 
the condensate function $F^{R(A)}$ (or $u^{R(A)}$) must equal zero when $x=1$. 
The solution is 

\begin{equation}
\hat{F}^{R(A)}\approx i \hat{\sigma}_y u^{R(A)}
= \frac{\hat{F}^{R(A)}_s \sinh (k^{R(A)}(1-x))}{r_b(k\cosh k)^{R(A)}}
\label{eq5}
\end{equation}

 The proximity effect is weak provided that $r_b \sqrt{\Gamma} >>1$ (a maximal 
value of $F^{R(A)}$ is achieved at $\epsilon=\Delta$). In this case the function 
$t(\epsilon)$ can be presented in the form \cite{r6}:$t(\epsilon)= 1/4<(F^{R}-F^
{A})^2>$. Performing the spatial averaging we obtain in the limit $l_1<<1$, 

\begin{equation}
\begin{array}{cc}
t(\epsilon)=(2r_b)^{-2} (\frac{|F_s|^2}{|k \cosh k|^2}
[\frac{\sinh 2k_1}{2k_1}-\frac{\sin 2k_2}{2k_2}]\\
+ Re (\frac{F_s^2}{(k \cosh k)^2}[\frac{\sinh (2k)}{2k}-1])\cos^2(\phi/2)
\end{array}
\label{eq6}
\end{equation}

where $k=k_1 + i k_2$.  
The first term in Eq. (\ref{eq6}) represents the anomalous MT contribution to 
the conductance variation and the second regular term is due to a DOS 
variation of the N wire caused by the proximity effect. One can easily check 
that the $partial$ conductance variation $t(\epsilon)$ is zero at $\epsilon=0$ 
(in this case $F^R = F^A$), increases as  $\epsilon^2$, 
reaching the first maximum when $\epsilon$ is of order max[1,$\gamma$] then 
decreases. At $\epsilon \approx \Delta$ the function $t(\epsilon)$ has a second 
maximum (see solid line in Fig. 2). With the aid of Eq.(\ref{eq1}) and (\ref{eq6}) we find 
the asymptotics of $\delta S$ at low temperatures $T<< \Delta$ (or $\beta \Delta 
>>1$). One has for the zero-bias $\delta S$ at low temperatures,

\begin{equation}
\delta S= \frac{\cos^2(\phi/2)}{(2r_{b})^{2}} c_0 \left\{
\begin{array}{cc}
 c_{1} \beta^{-2} , & \beta >> 1, (\gamma=0)\\
 5/\gamma^{7/2}\beta^{2} , & \gamma^{-1}<<\beta, 
(\gamma>>1)
\end{array}
\right.
\label{eq7}
\end{equation}
and at higher temperatures,
\begin{equation}
\delta S= \frac{\cos^2(\phi/2)}{(2r_{b})^{2}}\beta \left\{
\begin{array}{cc}
c_{2} , & \beta << 1, (\gamma=0)\\
c_{3} \gamma^{-1/2} , & \beta<<\gamma^{-1}, (\gamma>>1)
\end{array}
\right.
\label{eq8}
\end{equation}
 
 where the coefficients are, $c_0 \approx 0.82, c_1 \approx 0.86,
c_2 \approx 1.57$ and $c_3 \approx 1.56$. It is clear from the expressions 
(\ref{eq7} - \ref{eq8}) for $\delta S$
 that the conductance variation  $\delta S$
has a first maximum at a temperature 
$T_1 \approx max[\epsilon_{Th},\gamma]$ (we assume that both 
$\epsilon_{Th}$ and $\gamma$ are smaller than the zero temperature energy gap 
$\Delta(0)$).

 Let us turn to the case of high temperatures $T \approx \Delta$ when 
$\beta$ is close to the critical value $\beta_c=3.5/4\Delta(0)$.
 The contribution to $\delta S$ caused by a variation in the DOS; 
calculated by summing over the Matsubara frequencies, 
is small and of the order $\delta S \approx r_{b}^{-2} 
\beta^{3/2} (\Delta \beta)^2$. The main contribution 
is due to the MT term which can be presented in the form, 

\begin{equation}
\delta S_{MT} \approx  \beta_c(2r_b)^{-2} \int^{\infty}_{0}
d \epsilon \frac{\Delta^2}{\sqrt{(\epsilon^2 -\Delta^2 )^2+(2\epsilon \Gamma)^2}}
\chi(\epsilon)
\label{eq9}
\end{equation}
Here and in what follows, we set $\phi =0$ for brevity. Where
$\chi(\epsilon)=|k \cosh k |^{-2} [ \sinh (2 k_1)/2 k_1 -
\sin (2 k_2)/2 k_2]$. We replaced $\beta$ in Eq. (\ref{eq1}) by $\beta_c$ 
as $\beta$ depends on $\Delta$ very weakly, $\beta \approx \beta_c
[1+0.33(\Delta/\Delta(0))^2]$ where the second term is very small 
(the calculations were carried out for $\Delta(0)=20$). For small energies 
($\epsilon << max[1,\gamma]$) the function $\chi(\epsilon)$ is a constant 
equal to (2/3) when $\gamma<<1$ and equal to ($\gamma^{-3/2}$) when $\gamma>>1$, 
then for $\epsilon  >> max[1,\gamma]$ $\chi(\epsilon)$ decays as 
$\epsilon^{-3/2}$. The main contribution to $\delta S_{MT}$ stems from the 
singular region $\epsilon \approx \Delta$, therefore in the main 
logarithmic approximation we have $\delta S_{MT} \approx 1/2 \beta_c r_b^{-2} 
\Delta \ln(\Delta/\Gamma) \chi(\Delta)$ assuming that $(\Delta/\Gamma)>>1$. 
One can see that $\delta S_{MT}$ increases with increasing $\Delta$ from zero, 
reaches a maximum and then decreases.

 In Fig.3 we present the dependence $\delta S(T)$ calculated with the aid 
of Eqs (\ref{eq1}) and (\ref{eq6}). We see that besides the main peak at
$T_1 \approx 0.1T_c$ (i.e. $\tilde{T}_1 \approx \epsilon_{Th}$), there is another, 
weaker peak in the conductance near $T_c$. The temperature $T_2$ at which the 
second peak is achieved corresponds to the condition $\tilde{\Delta} \approx 7.6 
\epsilon_{Th}$. As the depairing rate $\gamma$ increases the first maximum is 
shifted towards higher temperatures, as predicted above.

  Fig. 4 shows the dependence $\delta S(\beta)$ for the same cross geometry 
in the case of a strong proximity effect ($r_b=0$), calculated 
using Eq. (\ref{eq3}). We see that the weak peak near $T_c$ has disappeared and 
the position and height of the first peak depends on $\gamma$ essentially as 
before, moving to higher temperatures with increasing $\gamma$.

 In Fig. 5 the temperature dependence of the conductance variation $\delta S 
(\beta)$ is shown for the structure in Fig. 1A. We assumed the weak proximity 
effect ($r_{b}=5$). The curves are presented for different resistances of 
the $N/N'$ interface ($r_{N/N'}$ is the ratio of the $N/N'$ interface 
resistance to the resistance of the N film). One can see that $\delta S$ can 
be negative. The negative sign of $\delta S$ is due to the shunting effect 
of the S strip which is stronger in the normal state (in the superconducting 
state the S/N interface resistance increases) \cite{r23}. As $r_{N/N'}$ 
increases, the 
height of the first maximum increases as the proximity effect is enhanced 
(the amplitude of the condensate function at the $N/N'$ interface is not zero 
if $r_{N/N'} \neq 0$ and increases with increasing $r_{N/N'}$). The second peak 
in $\delta S(\beta)$ becomes weaker and disappears at $r_{N/N'}=10$. This is in 
agreement with Fig. 4 where the strong proximity effect was considered 
($r_{b}=0$). It is worth noting that in this geometry the second maximum 
is more pronounced (if $r_{N/N'}$ is not large) than in the geometry of 
Fig.1B .In Fig. 6 we show the dependence of the second peak on the 
damping rate $\Gamma$ in the superconductor, as expected the second peak 
becomes less pronounced as $\Gamma$ increases.

 In Fig.2 we also plot the voltage dependence of the zero-temperature
conductance variation $\delta S(V)$ for the system shown in
 Fig.1B in which the N film is
replaced by a ferromagnetic film (F). We assume a weak proximity 
effect at very low temperature; in this case $\delta S(V) 
\approx t(\epsilon)$ with 
$\epsilon=eV$. Fig 2 shows that the dependence $\delta S(V)$ 
 with increasing exchange h (measured in units
$\epsilon_{Th}$) changes drastically. First, the zero-temperature 
$\delta S_0$ is not zero at zero bias (time reversal symmetry is broken)
and has a non-monotonic behaviour as a function of h
. Secondly, the low temperature peak 
is split and $\delta S$ approaches zero at a $V \approx h$ if 
$h>>\epsilon_{Th}$. We note that these effects can be observed only 
in the case of a weak ferromagnetism when $h<\hspace*{-0.4cm}_{\sim} 
\Delta$ (see for example 
Ref \cite{rr1} and references therein).

 In conclusion we have analysed the temperature dependence of the conductance 
$S(\beta)$ for mesoscopic $S/N$ structures of different geometries, and 
established that in the case of a weak proximity effect there are two peaks 
on the temperature dependence of $\delta S(T)$. The first maximum, as 
predicted earlier 
\cite{r4,r5,r6,r7}, corresponds to the temperature $T_1$ of order of the 
Thouless 
energy, and the second corresponds to the temperature 
$T_2$ at which $\Delta(T_2)$ is of order $\epsilon_{Th}$. Experimentally, this 
second maximum may already have been observed in \cite{r9}(a) where a drop 
in the resistance R was observed near $T_c$ followed by a smooth increase in 
$R$ with decreasing temperature ($400 mK<T< 1200 mK$). In contrast the 
first maximum 
occurs at a much lower temperature ($T_1 \approx 50 - 100 mK$).  
According to the theory presented above the position of the first maximum 
strongly depends on the depairing rate $\gamma$ in the normal wire, if 
$\gamma>\epsilon_{Th}$.

  A. F. Volkov is grateful to the Royal Society, to the 
Russian grant on superconductivity (Project 96053) 
and to CRDF (project RP1-165) for financial support.

\begin{figure}
\centerline{\psfig{figure=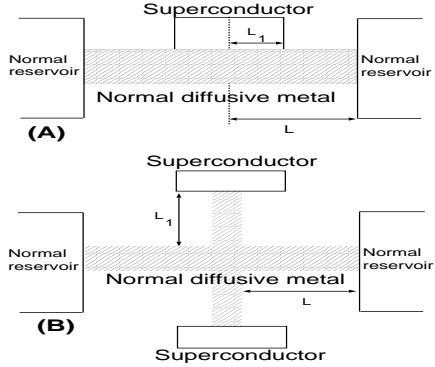,width=6cm,height=5cm}}
\caption{The structures considered.}
\label{fig1}
\end{figure}

\begin{figure}
\centerline{\psfig{figure=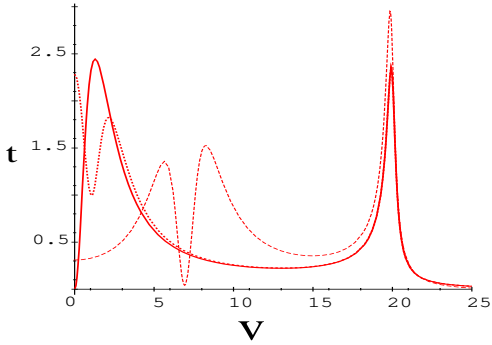,width=7cm,height=6cm}}
\caption{The dependence of $t(\epsilon)(2r_b)^2$ on energy, of 
the structure shown in Fig 1.(B), for different values of 
the exchange field $h=0$ (solid line)
, $h=1$ (thick dotted line) and $h=7$ (light dashed line),
With $\gamma=0.1$, $\Delta=20$,$\Gamma=0.3$,$l_1=0$. }
\label{fig2}
\end{figure}

\begin{figure}
\centerline{\psfig{figure=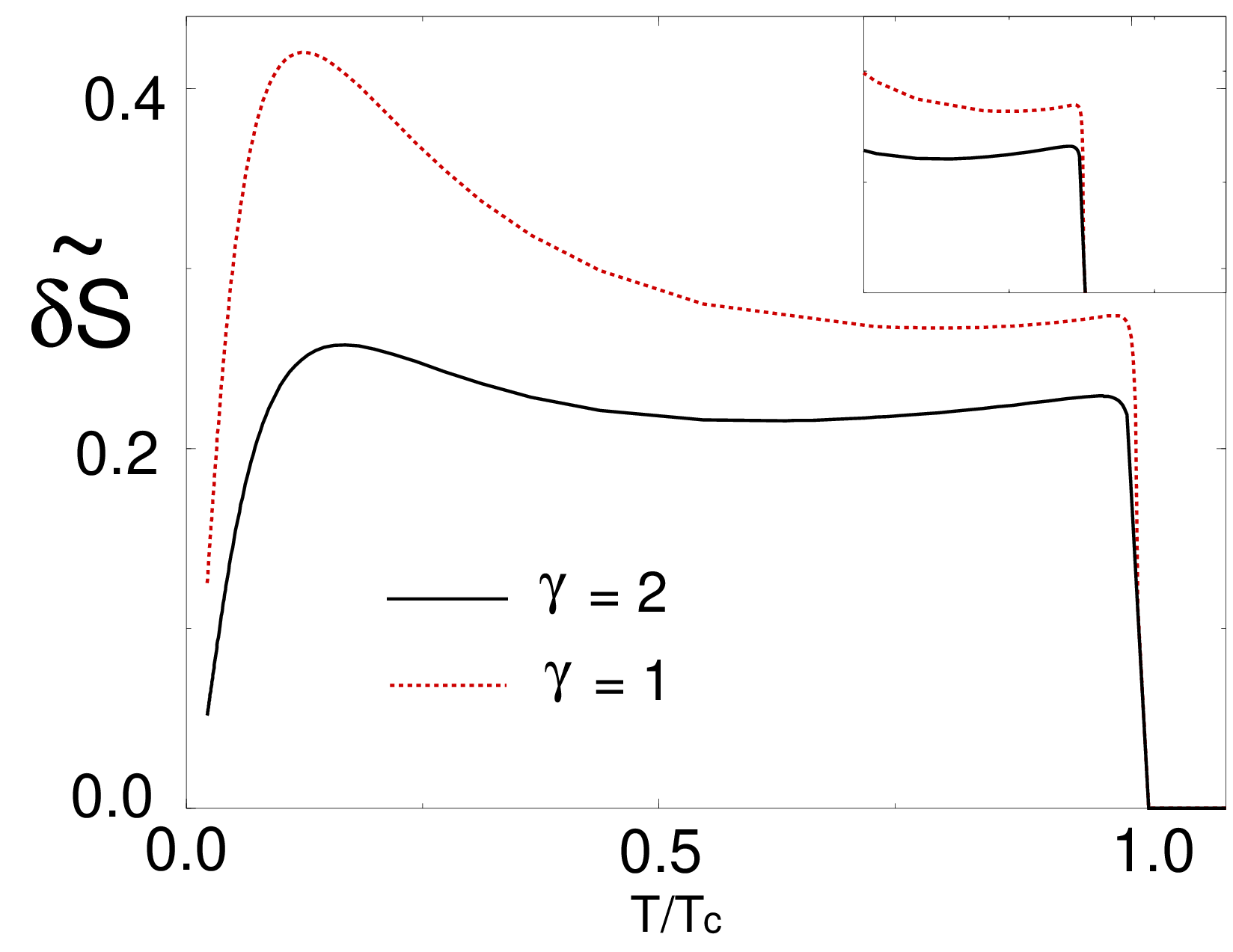,width=5cm,height=4cm}}
\caption{Dependence of $\delta \tilde{S}$ on temperature, for the cross
 geomentry in the 
weak proximity limit, for different depairing rates $\gamma$. With  
$\Delta_{0}=20$,$\Gamma=0.1$,$l_1=0$. The inset is an enlargement of the 
second maximum around $T/T_c=1.0$. Note the $\delta 
\tilde{S}=\delta S(2r_b)^2$.}
\label{fig3}
\end{figure}

\begin{figure}
\centerline{\psfig{figure=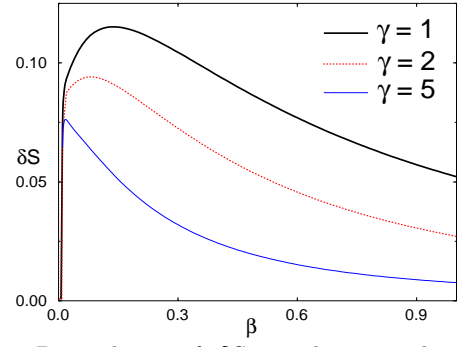,width=6cm,height=4.55cm}}
\caption{Dependence of $\delta S$ on the normalised inverse temperature, 
for the cross geomentry in the 
strong proximity limit. For different depairing rates $\gamma$. With  
$\Delta_{0}=100$,$\Gamma=0.1$,$l_1=0$.}
\label{fig4}
\end{figure}

\begin{figure}
\centerline{\psfig{figure=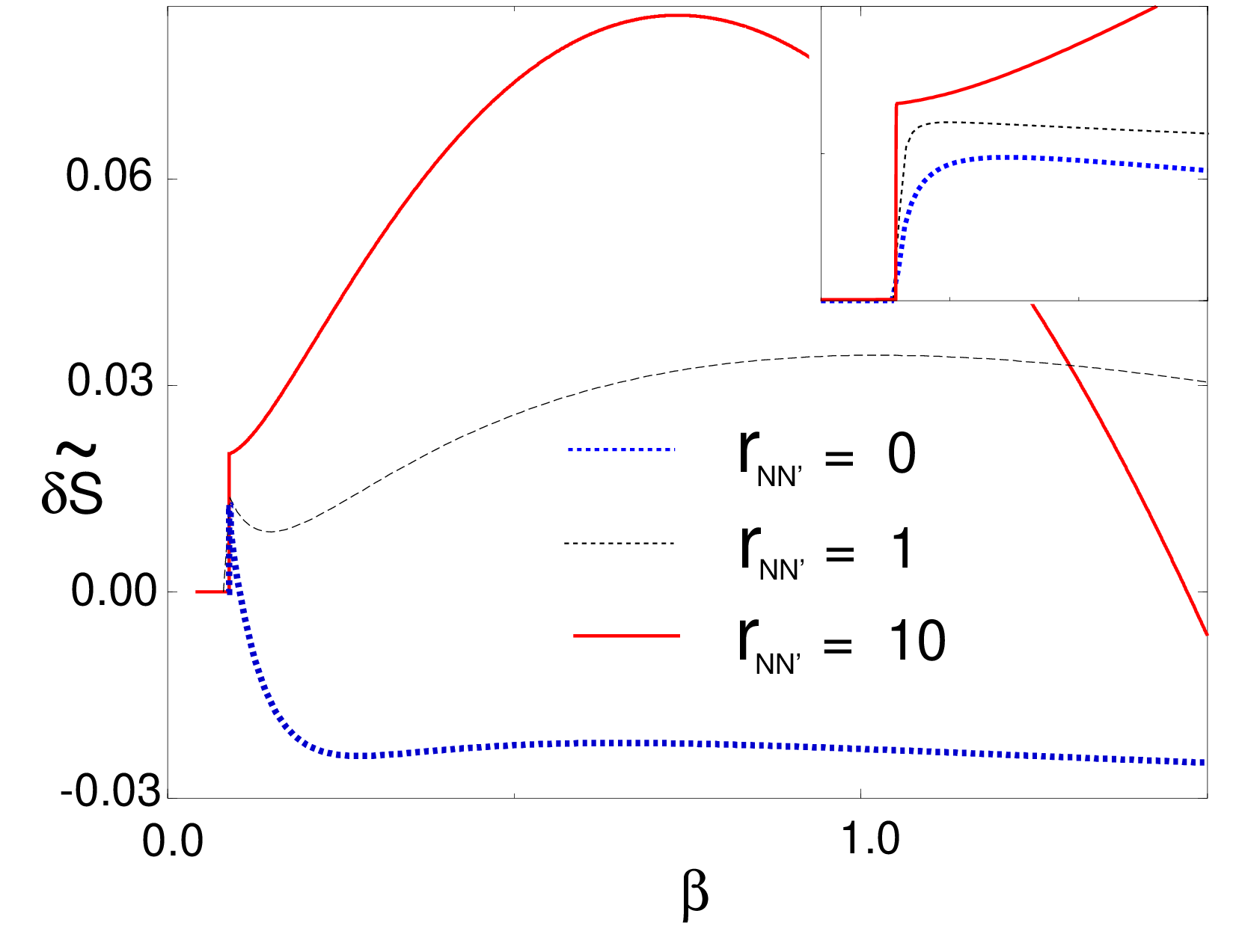,width=6cm,height=5cm}}
\caption{Dependence of $\delta \tilde{S} $ on the normalised inverse
 temperature, for 
the  geomentry shown in 
Fig. 1A in the weak proximity limit. For different interface 
resistances $r_{NN'}$
. With $\Delta_{0}=10$, $\gamma=0.1$,$\Gamma=0.1$,$l_1=0.2$. The inset is an
 enlargement of the 
second maximum around $\beta \approx 0.1$. Note $\delta \tilde{S}$ has been 
scaled for convenience, by 0.1 for $r_{NN'}=10$,0.2 for $r_{NN'}=1$ and 2 for 
$r_{NN'}=0$.}
\label{fig5}
\end{figure}

\begin{figure}
\centerline{\psfig{figure=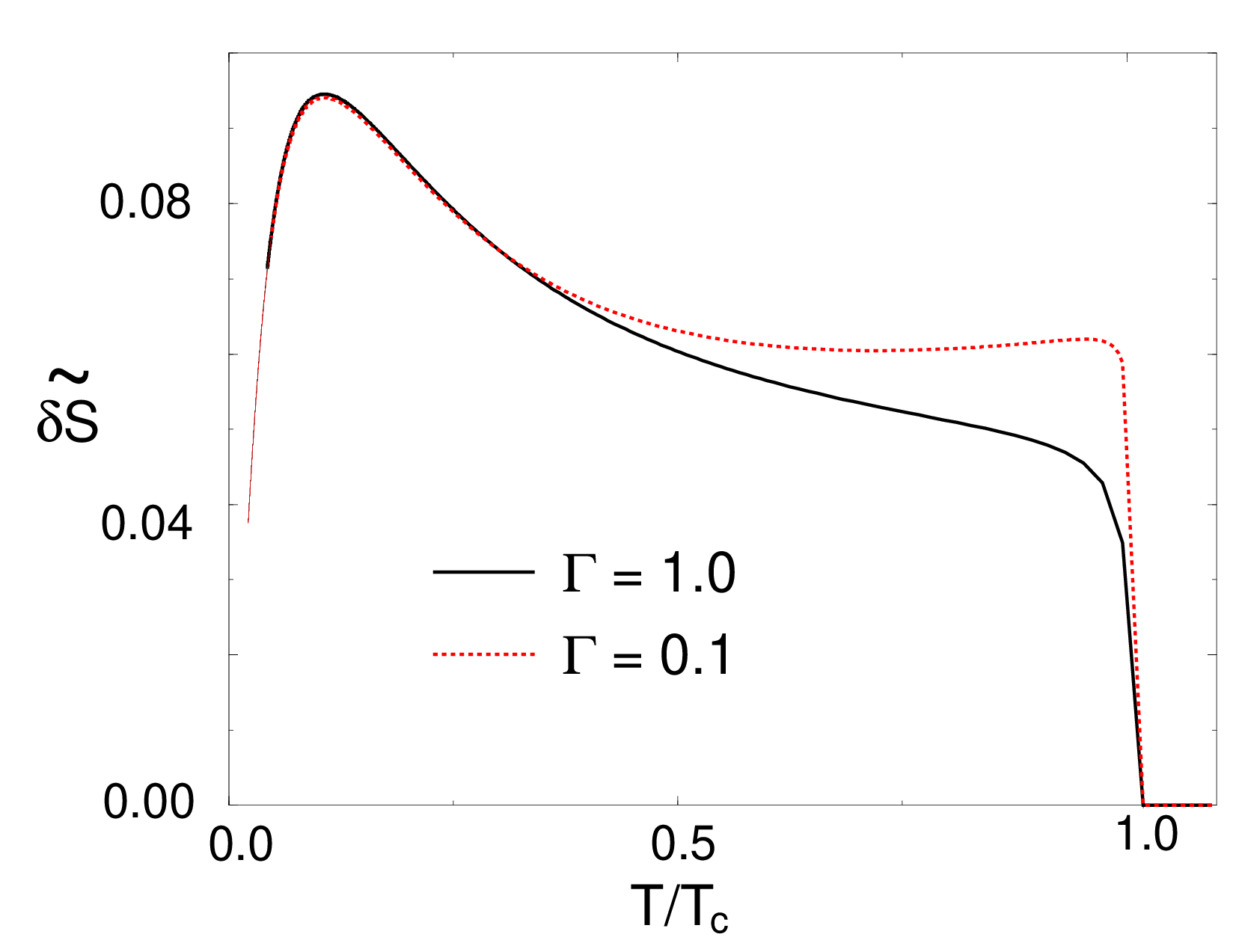,width=6cm,height=5cm}}
\caption{Dependence of $\delta \tilde{S}$ on temperature, for the  
geomentry shown in 
Fig. 1B in the weak proximity limit. For different damping rates $\Gamma$. 
With $\Delta_{0}=20$, $\gamma=2$,$\Gamma=0.1$,$l_1=0$.}
\label{fig6}
\end{figure}

\end{document}